\PassOptionsToPackage{table}{xcolor}
\documentclass[conference,compsoc]{IEEEtran}
\ifCLASSOPTIONcompsoc
  \usepackage[nocompress]{cite}
\else
  \usepackage{cite}
\fi
\usepackage[hidelinks]{hyperref}

\usepackage[most]{tcolorbox}

\definecolor{takeawaygreen}{RGB}{46,125,80}
\definecolor{takeawayback}{RGB}{244,250,246}

\newtcolorbox{takeawaybox}{
  enhanced,
  colback=takeawayback,
  colframe=takeawaygreen,
  boxrule=0.5pt,
  arc=2pt,
  left=6pt,
  right=6pt,
  top=5pt,
  bottom=5pt,
  before skip=6pt,
  after skip=6pt,
}
\usepackage{tikz}
\usepackage{booktabs}  
\usepackage{tabularx}  
\usepackage{multirow}  
\usepackage{amsmath}   
\usepackage{amssymb}   
\usepackage{dsfont}
\usepackage{amsfonts}
\usepackage{amsthm}
\usepackage{enumitem}
\usepackage[caption=false,font=footnotesize]{subfig}
\usepackage[table]{xcolor}
\usepackage{array}
\usepackage{tabularx}
\usepackage{cancel}
\usepackage{hyperref}

\usepackage{graphicx}
\usepackage{amsmath}
\usepackage{amssymb}
\usepackage[all]{nowidow}
\usepackage{siunitx}
\usepackage{multirow}
\usepackage{amsmath}
\usepackage{enumitem}
\usepackage{tcolorbox}
\usepackage[linesnumbered,ruled,vlined]{algorithm2e}
\usepackage{colortbl}
\usepackage{diagbox, eqparbox, hhline}
\usepackage{hyperref}
\usepackage{booktabs}

\usepackage{booktabs}
\usepackage{tabularx}
\usepackage{multirow}
\usepackage{makecell}
\usepackage{listings}
\usepackage{xurl}

\SetCommentSty{mycommfont}

\usepackage{wasysym}

\usepackage{xspace}

\makeatother

\newlist{rqlist}{enumerate}{3}
\setlist[rqlist]{label=\textbf{RQ\arabic*}.,before=\raggedright,leftmargin=40pt,ref=\textbf{RQ\arabic*}}

\newcounter{result}

\newcommand{\shortsection}[2][.]{\noindent\textbf{#2#1}}

\begin{document}
%
\title{The Limits of Information-Flow Control Enforcement for Agentic Systems}


\author{
\IEEEauthorblockN{Jonathan Gregory, Kyle Domico, Kunyang Li, Jean-Charles Noirot Ferrand, and Patrick McDaniel}
\IEEEauthorblockA{
University of Wisconsin--Madison, Madison, WI, USA\\
\texttt{\{jgregory8, domico, kli253, jcnf, mcdaniel\}@cs.wisc.edu}
}
}


%


\maketitle

\begin{abstract}

Information-flow control (IFC) is a promising approach to enforce policy in agentic AI systems. However, IFC suffers from label creep, a phenomenon where monotonically increasing security labels propagate through a system and progressively limit its utility. Label creep in agentic AI systems can be mitigated using two major strategies: (1) improved system design and (2) label declassification. In this paper, we characterize how far system design alone can mitigate label creep in agentic systems. Specifically, we measure how system topology, scale, and agent composition affect utility loss and security enforcement cost proxies for  conservative  IFC enforcement in a stateful agentic system simulator. We develop a pipeline to systematically vary these features and measure their impacts on label propagation. Our results demonstrate strong associations between structural graph features and our IFC utility loss metric (up to a Spearman correlation $|\rho|$ of $0.96$) and show that system size and tool allocation substantially affect how labels propagate. These findings identify design choices to reduce IFC utility loss, but they also demonstrate that system design alone cannot fully prevent utility loss due to monotonic label propagation. This motivates the development of label declassification mechanisms that aid these design choices and diminish IFC utility cost. 
\end{abstract}


%
\IEEEpeerreviewmaketitle

\section{Introduction}
\noindent Advances in the capabilities of large language models (LLMs) have led to the emergence of agentic computing, a paradigm in which AI agents---LLMs augmented with tools and persistent memory---execute tasks within a system. Working alone or together, these agents perform useful tasks that require autonomous problem-solving. However, they can also be very harmful, as evidenced by the recent 2026 OpenAI-HuggingFace incident, where a collection of OpenAI's agents performed unintended actions, communicated via shared channels, and exploited a vulnerability to infiltrate HuggingFace's infrastructure~\cite{openai_huggingface_incident_2026}.

The rapid adoption of AI agents to solve real-world tasks calls for stronger control mechanisms, especially for sharing information and using consequential tools. Information-flow control (IFC) provides a promising solution to this problem by controlling how security-labeled information flows through a system. In an agentic system, this could be used to authorize information sharing or prevent tool calls based on the security labels of the data accessed. However, to apply IFC efficiently to agentic systems, we need to better understand the challenges of doing so, specifically with respect to a phenomenon known as \textit{label creep}, where once information is marked as sensitive, every operation that it contributes to will also output information marked sensitive~\cite{sabelfeldLanguagebasedInformationflowSecurity2003a}. This can harm the utility of the system, as the IFC policy will likely block many (possibly benign) operations on this sensitive data. 

Traditional IFC systems reduce this utility loss through precise static tracking of information flows~\cite{sabelfeldLanguagebasedInformationflowSecurity2003a, myersJFlow1999}, restructuring computations into smaller security compartments~\cite{vassenaFromFine2019}, and controlled declassification mechanisms that safely relax security labels~\cite{myersComplete1998}. Precisely tracking information flows through LLM-generated outputs remains challenging in agentic systems~\cite{siddiquiPermissiveInformationFlowAnalysis2026, costaSecuringAIAgents2025}. This motivates two approaches for reducing label creep in agentic systems: modifying system structure to improve context isolation and developing mechanisms to declassify information. We focus on the former and ask: \textit{how much improvement can system design provide to IFC utility before declassification becomes necessary?}



In this work, we investigate how the structure (arrangement of agents), size (number of agents), and composition (tool calls assigned to agents) of agentic AI systems influence IFC security enforcement cost and utility loss. We focus on these components because administrators have direct control over them, making our findings actionable. By abstracting these elements, we perform a systematic analysis on how varying these components can impact security and utility costs under certain tasks. From this analysis, we characterize the security-utility tradeoffs of IFC enforcement.

We model agent tasks and system topologies as graphs from which we derive features. We then design a pipeline capable of generating task configurations that cover a variety of possible scenarios of different complexities. We motivate the problem by demonstrating how excess IFC utility loss worsens with increasing task complexity, which we model as the number of tool calls used to complete a task. Holding the task size constant, we then measure how system topology, size, and composition affect security enforcement cost, modeled by the number of communication vectors that must be monitored, and utility loss, modeled by the fraction of tool calls that introduce or are executed with a high-sensitivity label. Our analysis spans 4 standard network topologies, 5 random topologies of different densities, over 9,000 abstracted tasks composed of a variable number of interdependent tool calls with different orderings, and 2 tool clustering paradigms. Our pipeline runs independently of any LLM, enabling fast analyses of how different agentic system structures impact IFC security and utility.

Our analysis reveals that certain graph features are highly correlated with IFC utility cost (up to Spearman correlation $|\rho|=0.96$) and that system density is inversely related to utility loss. We also show that using more agents in a system lessens label propagation, but at the cost of up to quadratic growth in the number of edges that must be monitored. Finally, we show that isolating sensitive tool calls to specific agents can be effective for reducing overall IFC utility loss. Our analysis shows that structural design choices can reduce utility loss due to label propagation, but they cannot completely prevent it, motivating the development of  declassification mechanisms. 

\shortsection{Contributions} This paper provides three contributions. First, we develop an abstract, graph-based framework for isolating the impacts of system design on security and utility in a conservative IFC system. Second, we characterize how task complexity, inter-agent connectivity, system size, and sensitive tool allocation affect IFC enforcement cost and utility loss. Third, we motivate the development of declassification mechanisms by demonstrating that even under our ideal structural baseline, structural improvements alone cannot reliably prevent utility loss due to label propagation. 

\section{Background}
Information-flow control (IFC) is a method to ensure secure information flow by preventing leakage to forbidden outputs~\cite{denningLatticeModelSecure1976}. IFC can be used to protect confidentiality and integrity by preventing confidential or untrusted flows from reaching unauthorized or security-sensitive sinks\cite{myersProtectingPrivacyUsing2000}. IFC mechanisms commonly work by introducing a label---a tag indicating the information's confidentiality and integrity qualities---to data at a source and then propagating this label through the system with the data~\cite{enckTaintDroid2014}. IFC has been used to help secure agentic AI systems against prompt injection attacks~\cite{costaSecuringAIAgents2025, debenedettiDefeatingPromptInjections2025}, and prior work has investigated improving IFC utility for agents by hiding sensitive data~\cite{costaSecuringAIAgents2025} and only propagating labels from influential data~\cite{siddiquiPermissiveInformationFlowAnalysis2026}. However, these systems do not systematically examine the impact of agentic system design on IFC utility and enforcement costs. 

\section{Preliminaries}
\label{section:3}
\shortsection{System Abstraction}\label{subsection:system abstraction} Agentic systems involve highly complex tasks composed of interdependent tool calls executed by various agents. To abstract this, we model agentic systems as the following. Each topology is represented as a bidirectional graph $G=(V,E)$, where $V$ is the set of all agents in the system and $E$ is the set of bidirectional edges through which the agents can communicate. For a system composed of $n$ agents, $V=\{a_1, \ldots, a_n\}$ and $E\subseteq\{\{a_i, a_j\} : a_i, a_j \in V, i \neq j\}$. The exact configuration of edges in $E$ therefore dictates the topology. For the Star and Hierarchy topologies, additional routing agents are introduced (see \autoref{subsection:structure}). Next, let $T=\{t_1, \ldots t_m\}$ be the set of $m$ tool calls in a task, and let $T$ be allocated as $(T_{a_1}, \ldots, T_{a_n})$, which provides each agent with a set of tool calls such that no tool calls are shared between agents. Following \cite{yu2026benchmarking, shenTaskBench, kim2024an}, we model a task as a directed acyclic graph (DAG) $\tau = (T, D)$, where $D\subseteq T\times T$ is the set of directed tool dependencies, with $(t_i, t_j)\in D$ showing that $t_j$ depends on the output of $t_i$. Since $t_i$ and $t_j$ may belong to different agents, we denote these connections between agents with tool dependencies as $$C=\{\{a_i, a_j\} : i\neq j, \exists t_x \in  T_{a_{i}}, t_y \in T_{a_j} \ \text{with} \ (t_x, t_y) \in D\}.$$ 
    
\begin{figure}[t]
    \centering
    \includegraphics[width=\linewidth]{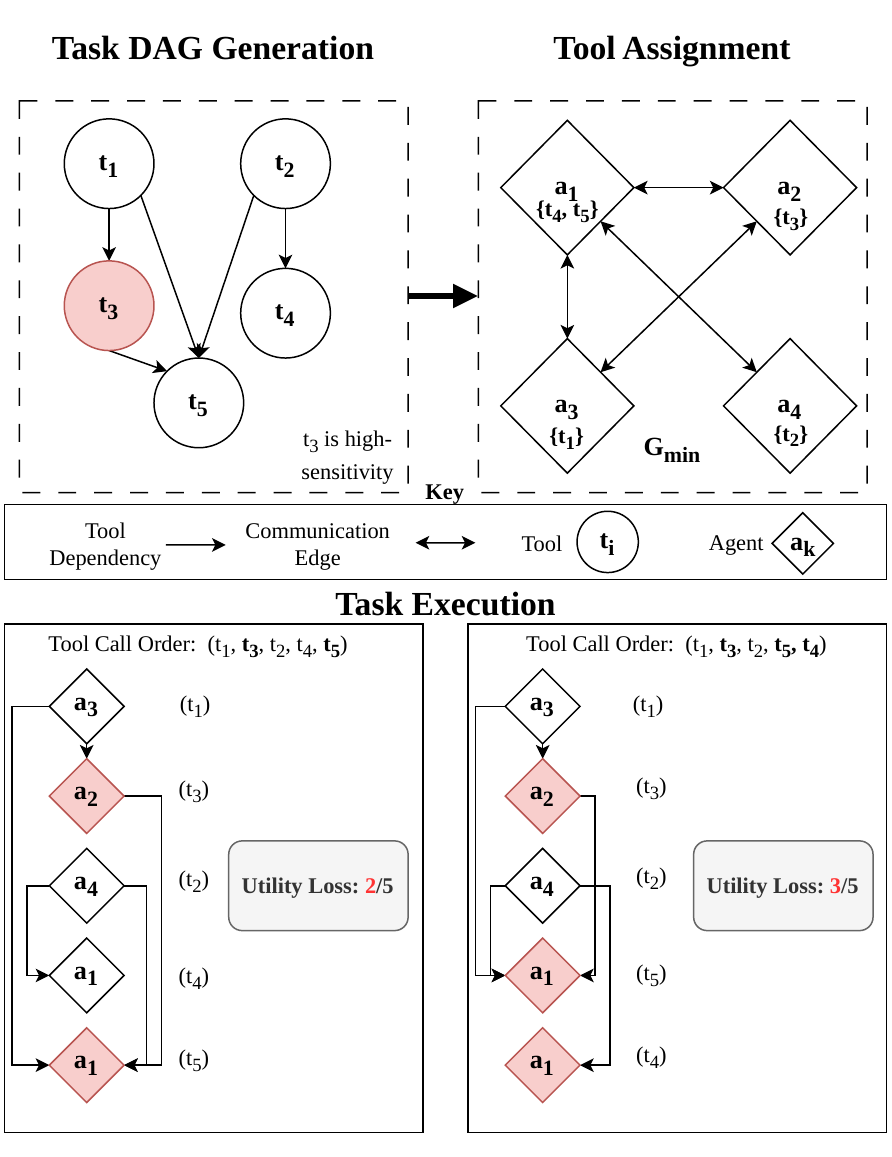}
    \caption{The task DAG, direct-communication baseline topology $G_{\min}$, and possible task executions resulting from different tool orderings.}
    \label{fig:abstraction}
    \end{figure}

    By including only the edges $C$, we obtain a graph containing the direct inter-agent connections induced by the task, which we call $G_{\min}=(V,C)$. $G_{\min}$ provides an ideal structural baseline for each DAG: every cross-agent tool dependency can be communicated directly without routing through intermediate agents, while the graph contains no communication edges not induced by the task.
    \autoref{fig:abstraction} shows an example of a task DAG, the corresponding $G_{\min}$ graph, and different tool call orderings for this task. 

    Depending on the actual topology, the edges in $G_{\min}$ may not exist in the actual graph. In these cases, the resulting paths needed to transmit context between agents may be longer. This is a primary motivation for why varying topology may impact label creep: it directly impacts how context is transmitted across graphs. The goal of an optimal topology is therefore to minimize the spread of context (and labels) for each system.  
    

\shortsection{Information-Flow Control Assumptions}
    We model our system's IFC behavior using the following assumptions. First, we assume that any tool $t$ can introduce a label $\ell$ into the system at the point where $t$ is called. Second, we assume that labels can propagate in two ways: as context to another tool through a mapped dependency or as context being routed between agents. Third, we assume that the IFC system operates on the basis of conservative joins---that is, if an agent with a label $\ell_l$ is exposed to a label $\ell_h$ and $\ell_l \preceq \ell_h$ (i.e., $\ell_h$ is at least as sensitive as $\ell_l$), the agent's new label is $\ell = \ell_l \sqcup \ell_h = \ell_h$. Fourth, we assume agents are stateful, so if an agent is exposed to context that contains a label---either through the result of a tool call or by routing labeled context to another agent---the agent will take the conservative join of that label with its previous label. Assuming stateful agents reflects the role of persistent memory in agents. Importantly, there is no mechanism in this system to declassify a label or prevent its transmission, leading to monotonically increasing label propagation. 

\section{Methodology}
Our goal is to measure security enforcement cost and utility loss in agentic systems as they execute a variety of complex tasks. To do so, we abstract our system topologies and tasks using the formulas described in \autoref{subsection:system abstraction}. For experiment reproducibility, we outline our methods as follows. First, in \autoref{subsection:structure}, we define the topologies used in this research, the task generation process, and label placement in tool calls. Then, in \autoref{subsection:measurement}, we define proxy metrics used to examine the graph structure, security cost, and utility loss of the system. 

\subsection{Structure}\label{subsection:structure}
\shortsection{Topologies}
We implement each system of $n$ tool-calling agents using four common graph topologies: Mesh, Star, Ring, and Hierarchy. In the Ring topology, order was determined via agent label (i.e., $[1,n]$). In the Star topology, an additional orchestrator agent was introduced as the hub. In the Hierarchy topology, agents were placed as leaves of a balanced tree with two main branches overseen by manager agents and a root managed by a supervisor agent. Only the $n$ tool-calling agents are assigned tool calls; the orchestrator, manager, and supervisor agents exist to route information in these topologies. We introduce six additional topologies to this system: the ideal topology for a task (specified by $G_{\min}$) and five random topologies (ER) generated by the Erdős-Rényi model with connection probabilities ($p\in \{0.1, 0.2, 0.3, 0.4, 0.5\}$) that continue sampling until the graph is connected and each graph is distinct.

\shortsection{Task Generation} \label{subsection:task-generation}
We generate arbitrary agentic tasks as random dependency graphs. For a task composed of $m$ tool calls, we first generate a random DAG using the Erdős-Rényi model with $p=\frac{2}{m+1}$, meaning that the expected number of dependencies will be around $m$. Next, we uniformly add each tool call in the task DAG to an agent's tool set. Finally, we use this resulting task DAG to build $G_{\min}$, which is the direct-communication baseline topology needed to complete the task. As a heuristic, we model task complexity as proportional to the number of tool calls used by a task. 

\shortsection{Label Placement}\label{subsection:labeling}
We simulate a system where calling certain tools introduces a high-sensitivity label into the system. It is then added to the running context of the agent at the point where the tool is called. In our experiments, we use the set of low- and high-sensitivity labels $\Lambda=\{\ell_l, \ell_h\}$, with $\ell_l \prec \ell_h$ and where $\ell_l \sqcup \ell_h = \ell_h$. We then use two different ways to introduce labels into the system via tool calls. The first is random placement: with a labeling budget of $b$, we randomly mark $b$ tool calls with an $\ell_h$ label and all others with an $\ell_l$ label. This simulates scenarios in which any agent could execute a tool call that can introduce sensitive information into the system. The second approach is clustering tool calls with $\ell_h$ labels to specific agents and assigning $\ell_l$ to the remaining tool calls for the task.  

\subsection{Measurement}\label{subsection:measurement}
\shortsection{Graph Features}
To compare agent topologies, we use the following features that capture graph interconnectedness and complexity. We use average shortest-path length, diameter, and density to measure graph structure and shape. We then use average betweenness centrality to measure routing through specific nodes in the graph and the clustering coefficient to measure how closely nodes are clustered in the graph. Precise definitions are given in~\autoref{tab:graph-metrics}. 

\shortsection{Security Metrics}
We use the number of communication edges as a proxy for security enforcement cost. This simulates how networks with fewer connections are simpler to monitor and have fewer connections that could allow threats to propagate in unexpected ways. For a graph $G=(V,E)$, 
\begin{equation*}
    \mathrm{S}(G) \propto |E|, 
\end{equation*}
where $\mathrm{S}$ is the security enforcement cost. 

\shortsection{Utility Metrics}
Label propagation is damaging to utility because it increases the risk that certain tool calls will be quarantined or denied if they introduce or occur in a high-sensitivity state. We measure label propagation as a proxy for utility loss via the formula 
\begin{align*}
    \mathcal{L}_{} (\tau, G)  = \frac{\text{\# of high-sensitivity tool calls}}{\text{\# total tool calls}},
\end{align*}
where $\mathcal{L}$ is the utility loss, $\tau$ is the task DAG, and $G$ is the agent graph. Some high-sensitivity execution is unavoidable under this conservative IFC configuration, but the amount of avoidable utility loss is measured as: 
\begin{align*}
    \mathcal{L}_{\text{excess}}(\tau, G) = \mathcal{L}_{}(\tau, G)-\mathcal{L}_{}(\tau, G_{\min}). 
\end{align*}
Here, $\mathcal{L}_{\text{excess}}$ tells us how much of the excess label propagation we see is a function of our specific topology, rather than the task, as $G_{\min}$ is the direct-communication baseline for $\tau$. In our experiments, we aim to find ways to minimize $\mathcal{L}$, but we also measure $\mathcal{L}_{\text{excess}}$ to determine how well a given topology performs in comparison to an ideal baseline. 
\section{Evaluation}
We aim to answer the following research questions: 

\begin{figure}[t] 
    \centering
    \includegraphics[width=\linewidth]{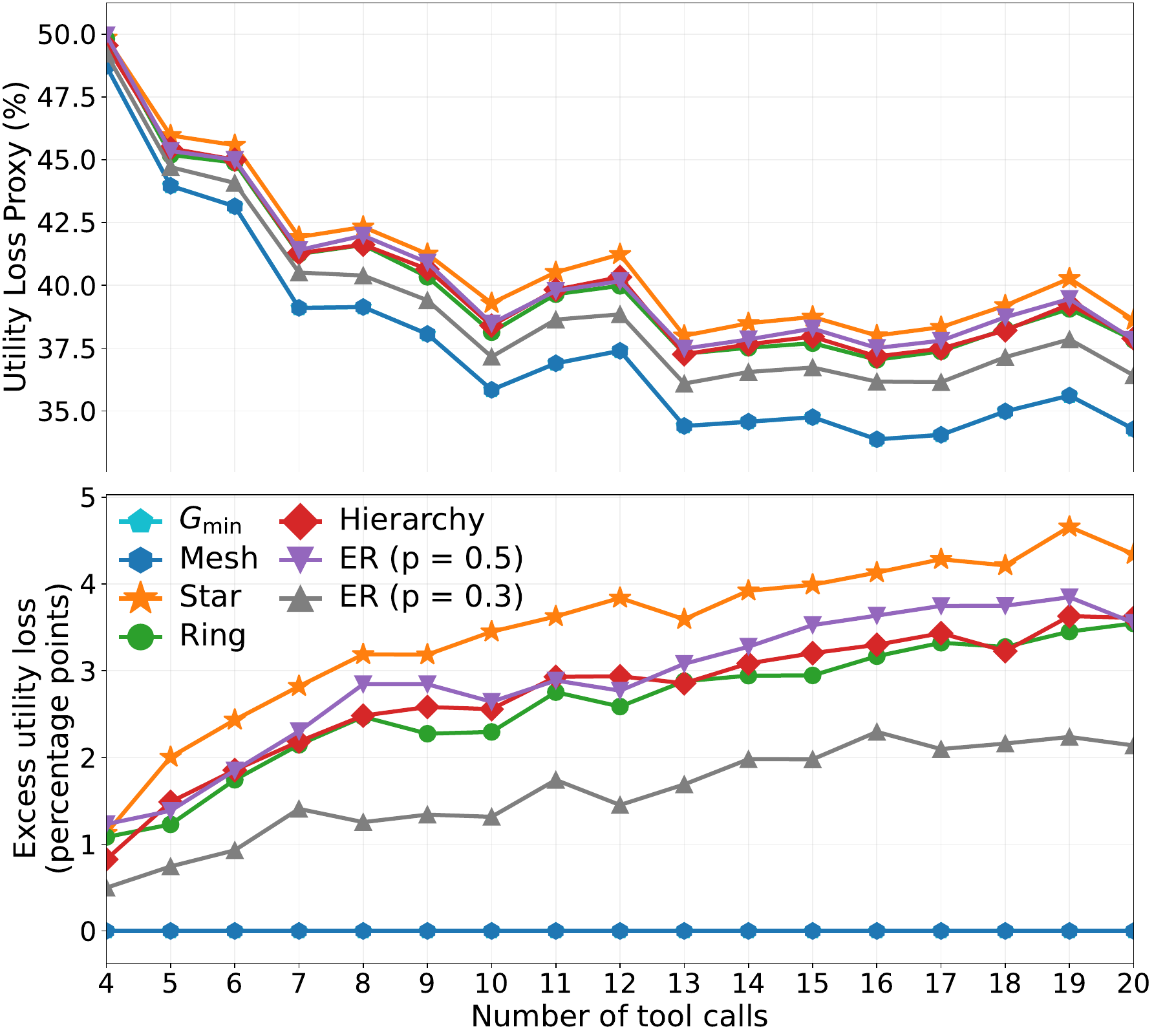}
    \caption{Gross utility loss and excess utility loss with increasing task complexity. Different topologies experience different levels of gross and excess utility loss, motivating this research. Mesh and $G_\text{min}$ overlap.}
    \label{fig:label-task-complexity}
\end{figure}

\begin{rqlist}[series=rquestion]
    \item How does task complexity affect IFC utility loss across agent topologies?\label{rq1}
    \item Which graph features are most correlated with IFC utility loss?\label{rq2}
    \item Does the number of agents in a topology affect IFC security enforcement cost and utility loss?\label{rq3}
    \item Does sensitive tool allocation affect IFC utility loss? \label{rq4}
\end{rqlist}

\subsection{Experimental Setup}
\label{subsection:sampling} We simulate task execution in each topology as follows. First, we run a topological sort on the task DAGs and process the tool calls in valid orders. Connected tool calls in the DAGs may also require routing between agents, which is always performed via the shortest path from one agent to the next. Next, tool calls that are not connected but are processed sequentially require a jump, either to the agent containing the tool call's immediate predecessor or, if no such dependency exists, directly to the agent containing the tool call itself. The exceptions to this are the Star and Hierarchy topologies: in cases where there is no direct predecessor to the tool call, the jump is to the orchestrator or supervisor, respectively. Jumps do not pass context, but direct communications do. This context passing occurs from all immediate predecessors on which the current tool call relies. The context delivery order can also affect label propagation, so we vary it with the topological orderings to fairly sample executions. The full topological orderings and delivery options for these DAGs can be very large, so where exhaustive enumeration is not feasible, 10,000 ordering-delivery samples for each DAG are randomly sampled and used for the analyses. For consistency, we use a set of 1,000 task DAGs containing 20 tool calls for all experiments, except where otherwise noted.

\subsection{Impact of Task Complexity}\label{subsection:complexity}
We begin with a motivating question (\ref{rq1}) for this research: \textit{How does task complexity affect IFC utility loss across agent topologies?} In other words, will increasing task complexity (modeled as the number of interdependent tool calls) impact the utility loss more in some systems than others? Answering this indicates whether system design is a reasonable experimental variable to affect IFC utility loss. 

To simulate increasing task complexity, we generate $500$ DAGs each for tasks ranging from 4 to 20 tool calls. We then randomly mark one tool in each DAG as sensitive, run each DAG against our topologies populated with 5 tool-calling agents, and plot the results, shown in \autoref{fig:label-task-complexity}. First, note that utility loss scales differently across topologies as task complexity increases. This observation motivates our central hypothesis that agentic system structure affects IFC utility. Second, excess utility loss increases with increasing task complexity. This indicates that as tasks use more interdependent tool calls, agent topologies that omit edges from $G_{\min}$ will diverge from the ideal label propagation rate.

\begin{tcolorbox}[colback=gray!5!white, colframe=gray!75!black]
  \textbf{\ref{rq1} Takeaway:} Task complexity does affect IFC utility loss, and increasing task complexity increased excess utility loss in non-baseline
agent topologies. 
\end{tcolorbox}

\subsection{Impact of Topology} \label{subsection:topology}
\begin{figure}[ht]
    \centering
\includegraphics[width=\linewidth]{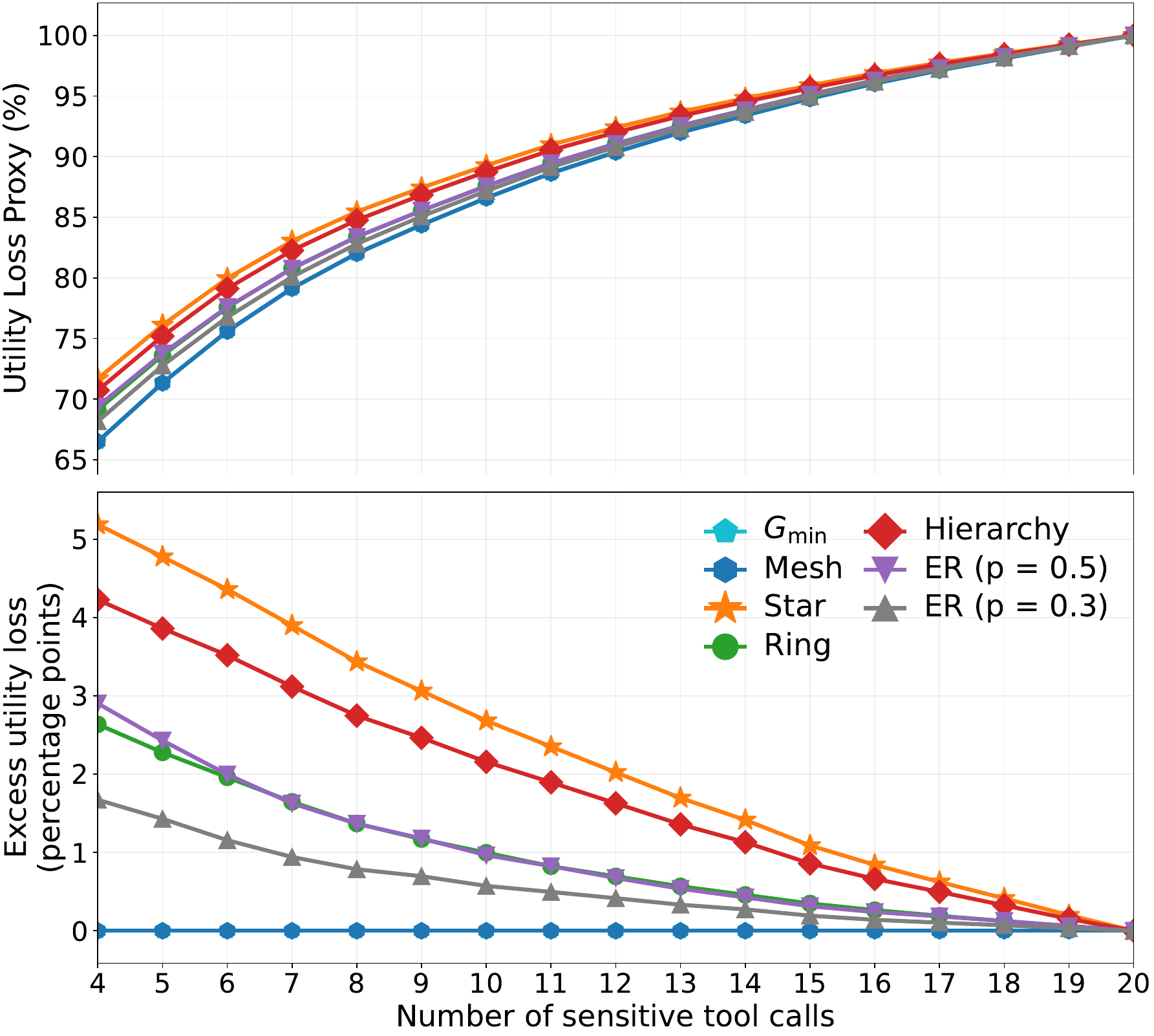}
    \caption{Label budget sweep across topologies.}
    \label{fig:sweep-topologies}
\end{figure}

From \autoref{subsection:complexity}, we see that graph topology does impact IFC utility loss. In this section, we examine why this occurs. In particular, we aim to answer \ref{rq2}: \textit{Which graph features are most correlated with IFC utility loss?} To answer this, we experiment with how different topologies experience different rates of label propagation in the presence of varying tool sensitivity. Using the methodology and task DAGs described in \autoref{subsection:sampling}, we generate 10,000,000 simulated task runs on each of the 10 topologies composed of 5 agents. We then conduct a sweep across labeling budgets using the random labeling strategy from \autoref{subsection:labeling} with $b\in [4, 20]$. \autoref{fig:sweep-topologies} presents the results of this sweep, with ER graphs with $p\in\{0.1, 0.2, 0.4\}$ omitted for clarity. As the figure shows, there is clear stratification between the topologies in terms of both utility loss and excess utility loss. For example, at $b=4$, excess utility loss is approximately 5 percentage points for Star, 4 for Hierarchy, and 3 for Ring. $G_{\min}$ and Mesh both perform the best in this respect, which is expected since these graphs encode direct connections between all agents that must communicate. The remaining topologies demonstrate an interplay between connectedness, complexity, and label propagation. 

\begin{figure}[h]
    \centering
    \includegraphics[width=\linewidth]{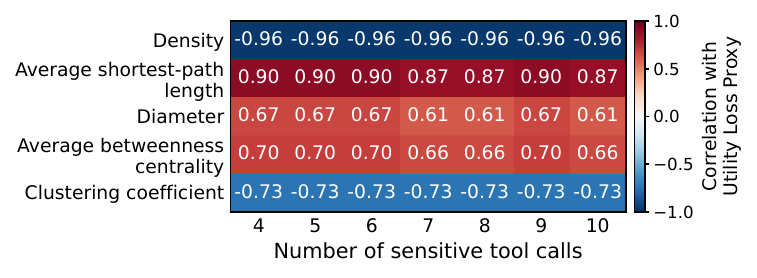}
    \caption{Spearman correlation $\rho$ between graph features and IFC utility loss for labeling budgets in [4, 10].}
    \label{fig:correlation_heatmap}
\end{figure}

To better understand these relationships, we compute Spearman correlations across the nine topologies, pairing each topology’s structural features with its mean utility loss over the simulated task executions. \autoref{fig:correlation_heatmap} shows these results. Since $G_{\min}$ is not a consistent topology (i.e., it varies based on the task), it is omitted from this analysis. 

There exists a strong negative association between utility loss and density ($\rho=-0.96$), which is logical since density measures the fraction of possible direct links present, which can minimize label propagation by limiting routing through intermediate nodes. Conversely, there is a high positive association between utility loss and average shortest-path length ($\rho=0.87$ at $b=10$). This makes sense, as a smaller average shortest-path length indicates more nodes are closely connected, which also minimizes routing. These statistics help explain surprising phenomena. For example, the Ring topology exhibited greater utility loss than the sampled $p=0.3$ ER graph, which was denser (0.7 vs. 0.5) and had a shorter average shortest-path length (1.3 vs. 1.5). By contrast, the sampled $p=0.5$ ER graph had the same density as Ring (0.5), a similar average shortest-path length (1.6), and similar utility loss. This correlation allows us to answer the research question.

\begin{tcolorbox}[colback=gray!5!white, colframe=gray!75!black]
  \textbf{\ref{rq2} Takeaway:} Interconnectedness metrics (especially density and average shortest-path length) are strongly correlated with IFC utility loss. 
\end{tcolorbox}

\begin{figure}[t]
    \centering
\includegraphics[width=\linewidth]{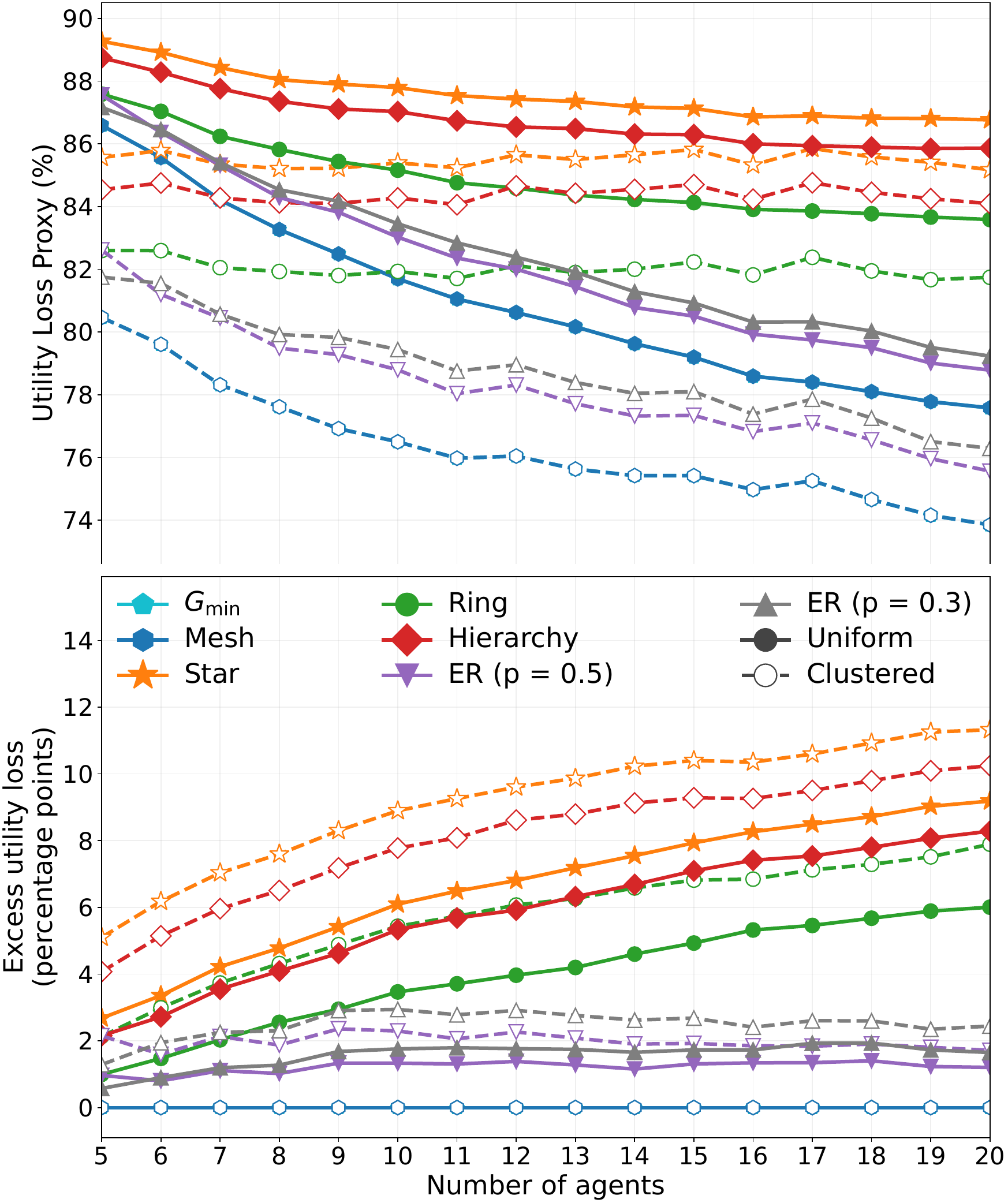}
    \caption{Gross and excess utility loss by number of tool-calling agents with a sensitive tool budget $b=10$. Solid lines show random allocation. Dashed lines show clustered allocation.}
    \label{fig:sweep-agents}
\end{figure}

\subsection{Impact of Agent Count} \label{subsection:number}
Next, we examine whether changing the number of agents in the topology impacts utility loss. Specifically, we examine \ref{rq3}: \textit{Does the number of agents in a topology affect IFC security enforcement cost and utility loss?} We use the same setup as in \autoref{subsection:topology}, but instead of fixing the system at 5 agents, we vary the number of agents from 5 to 20. As before, tool calls---labeled or not---are assigned with uniform probability to each agent up to a budget. First, we study utility loss with an original labeling budget $b=10$. \autoref{fig:sweep-agents} shows the results of these runs as solid lines. Observe that as we increase the number of agents, utility loss decreases while excess utility loss mostly increases for all topologies but the $G_\text{min}$, Mesh, and ER graphs. Besides $G_\text{min}$, these are also the types of topologies in which the number of edges increases quadratically with an increase in the number of agents. This suggests that the stronger utility improvement in these topologies is associated with their greater growth in direct connectivity.

\begin{figure}[ht] 
    \centering
    \includegraphics[width=\linewidth]{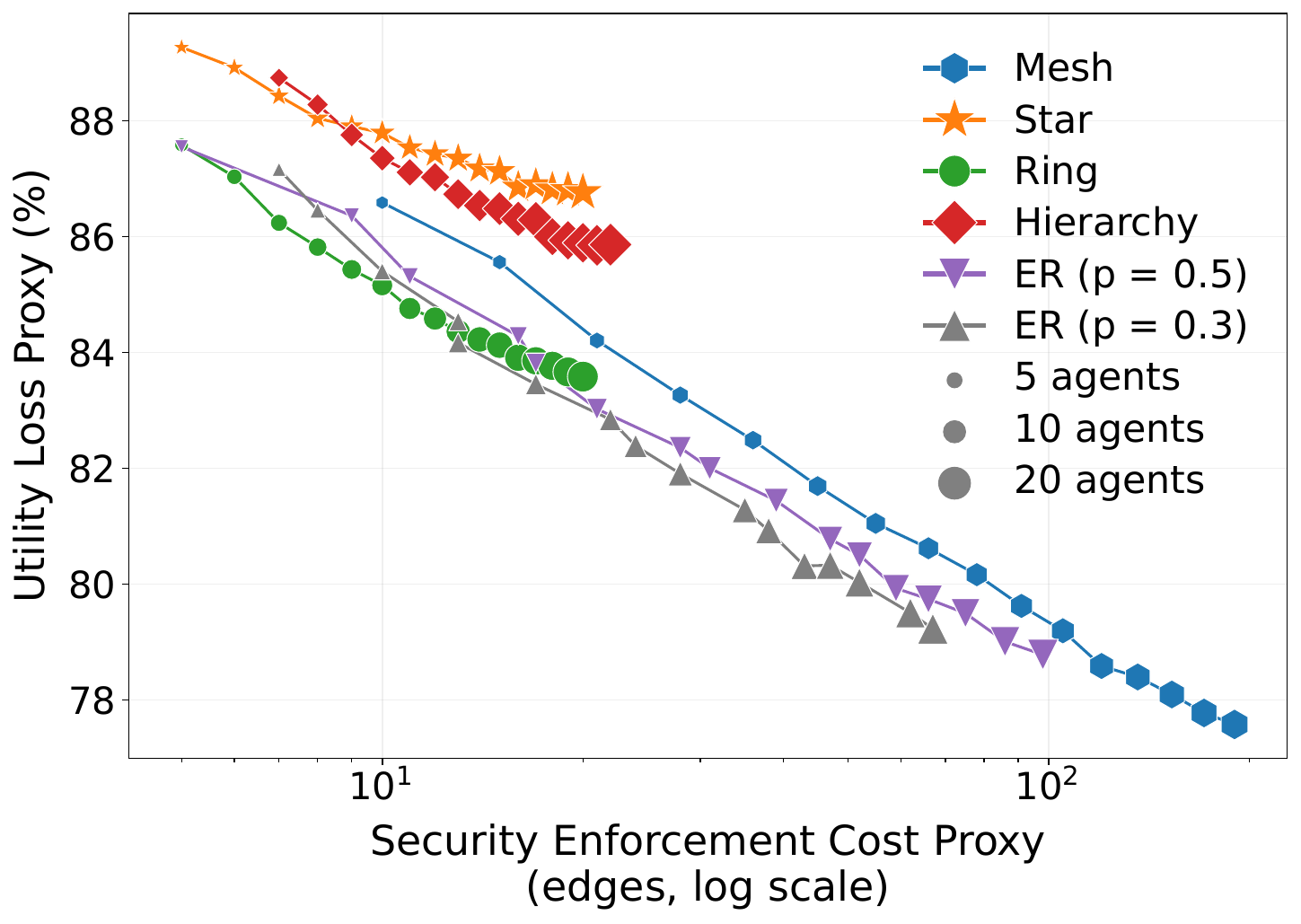}
    \caption{Utility loss proxy vs. security enforcement cost proxy for tasks with 10 sensitive calls. Larger markers indicate more tool-calling agents.}
    \label{fig:security-utility}
\end{figure}

However, as we increase the number of edges in the system, this also has a direct effect on our security enforcement cost. \autoref{fig:security-utility} shows this plainly, demonstrating that increasing the number of agents in the system leads to utility improvements at an increased security enforcement cost. These results illustrate a security-utility tradeoff of using system design to improve IFC enforcement under our model. 

\begin{tcolorbox}[colback=gray!5!white, colframe=gray!75!black]
  \textbf{\ref{rq3} Takeaway:} As the number of tool-calling agents increased, IFC utility loss decreased while the security enforcement cost proxy increased.
\end{tcolorbox}

\subsection{Impact of Tool Assignment}
Finally, we analyze \ref{rq4}: \textit{Does sensitive tool allocation affect IFC utility cost?} We examine whether isolating label-inducing tool calls to specific agents can minimize utility loss. In a real-world implementation, this requires assigning one agent with a set of high-sensitivity tool calls---for example, creating a dedicated banking agent that manages API calls to read and synthesize financial data. To perform this experiment, we repeat the tasks from \autoref{subsection:number}, but when assigning tool calls to agents, we design it so that labeled tool calls cluster together in specific agents' sets. The results are shown in \autoref{fig:sweep-agents} as dashed lines. 

Observe that although the clustered tool assignments follow a similar ordering in terms of best-to-worst utility loss as the random tool assignments, the utility loss statistics for the clustered tool calls uniformly beat their random counterparts. However, the clustered tool assignments also have higher excess utility cost statistics compared to their random counterparts, indicating that clustering sensitive tools improves the label propagation rates of the ideal graphs more than those of the non-ideal ones. 

\begin{tcolorbox}[colback=gray!5!white, colframe=gray!75!black]
  \textbf{\ref{rq4} Takeaway:} Sensitive tool call clustering lowered utility loss but increased excess utility loss for non-ideal topologies compared to random allocation.
\end{tcolorbox}

\section{Discussion}

\shortsection{System Design} Our experiments show that highly interconnected architectures are effective at combating label propagation. This is logical, as these architectures allow context to be routed directly from a source to a destination without touching any intermediate nodes. However, implementing highly interconnected structures is costly in terms of security enforcement. Our analysis shows that random networks that are sparser, such as the ER graph with $p=0.3$, can approach this performance without incurring the same security tradeoffs. For example, before conditioning on connectivity, the ER graph with $p=0.3$ requires $70\%$ fewer edges in expectation than a Mesh. Overall, our analysis indicates that by increasing interconnectedness in communication structure, we can create agent networks that resist IFC utility loss, but this must be balanced against security enforcement cost. Our experiments also suggest that increasing the number of agents in the system can reduce utility loss, as each of these agents' contexts serves as a form of isolation against label creep. Finally, our results indicate that isolating sensitive tool calls to specific agents can reduce utility loss. 

\shortsection{Future Work}
Our simulations isolate the effects of agentic system structure on IFC utility loss; future work will measure these trends in real agent workloads. However, regardless of the structure used, utility loss is pervasive. This indicates there is a limit to using system design alone to mitigate utility loss related to label propagation. In our model, this limit is specified by $G_{\min}$. As such, complementary mechanisms are needed to reduce utility loss beyond what is achievable through architecture alone. In future work, we will use lessons from this research to design declassification mechanisms like those in traditional IFC systems to safely declassify labels as they accumulate through the system. 
\section{Conclusion}
In this measurement study, we examined how topology, system size, agent composition, and tool allocation affect security enforcement cost and utility loss in agentic AI systems. We also characterized which graph features are strongly associated with utility loss and highlighted the security-utility tradeoffs inherent in designing more interconnected systems. This work indicates that information-flow control enforcement in agentic AI systems can be improved by using systems that balance direct communication and context isolation. However, this work shows that there is a limit to how much utility improvement is possible using system design optimization alone. Future work will examine how to use this insight to design effective declassification mechanisms that faithfully preserve system utility while maintaining security guarantees in agentic AI systems.

\ifCLASSOPTIONcompsoc
  \section*{Acknowledgments}
\else
  \section*{Acknowledgment}
\fi
The authors would like to thank Blaine Hoak and Yohan Beugin for editorial feedback and the anonymous reviewers for suggestions that helped clarify our results. This material is based upon work supported by PRISM, one of seven centers in JUMP 2.0, a Semiconductor Research Corporation (SRC) program sponsored by DARPA.

\section*{Artifact Availability}
The source code and experimental artifacts are available at 
\url{https://github.com/Grego3j/The-Limits-of-Information-Flow-Control-Enforcement-in-Agentic-Systems}.

\section*{LLM Usage Statement}
Generative AI was used to assist with the Python implementations of this research. LLMs were also used to conduct preliminary background research. Finally, LLMs were used to review the paper for grammatical and structural errors and suggest revisions. The research questions, experimental design, analysis, insights, and technical claims were developed by the authors.

\bibliographystyle{IEEEtran}
\bibliography{references}

\appendices
\section{Graph Features}
\begin{table}[h]
\centering
\caption{Graph Features for Agent Topologies}
\label{tab:graph-metrics}
\scriptsize
\setlength{\tabcolsep}{3pt}
\renewcommand{\arraystretch}{1.5}
\begin{tabularx}{\columnwidth}{
    @{}p{0.36\columnwidth}
    >{\raggedright\arraybackslash}X@{}
}
\toprule
\textbf{Metric} & \textbf{Formula} \\
\midrule
Average Shortest Path
& $\displaystyle \frac{2}{N(N-1)}
   \sum_{u<v} d(u,v)$ \\

Diameter
& $\displaystyle \max_{u<v} d(u,v)$ \\

Density
& $\displaystyle \frac{2M}{N(N-1)}$ \\

Betweenness Centrality 
& $\displaystyle b(v)=\frac{2}{(N-1)(N-2)}
   \sum_{\substack{s<t\\s,t\neq v}}
   \frac{\sigma_{st}(v)}{\sigma_{st}}$ \\

Average Betweenness Centrality 
& $\displaystyle \bar{b}=\frac{1}{N}\sum_{v\in V}b(v)$ \\

Clustering Coefficient
& $\displaystyle \mathcal{C}=\frac{1}{N}\sum_{v\in V }c(v)$,
  \quad
  $\displaystyle c(v)=
   \begin{cases}
   \frac{2T_v}{k_v(k_v-1)}, & k_v\geq 2,\\
   0, & k_v<2
   \end{cases}$ \\
   
\bottomrule
\end{tabularx}
\smallskip
\begin{minipage}{\columnwidth}
\scriptsize
$N=|V|$ and $M=|E|$;  
$d(u,v)$ is shortest-path length;
$\sigma_{st}$ counts shortest paths from $s$ to $t$, and
$\sigma_{st}(v)$ counts those paths passing through $v$, excluding endpoints; 
and $k_v$ is degree and $T_v$ counts triangles containing $v$.
\end{minipage}
\end{table}

\cleardoublepage



%



\end{document}